\documentclass[12pt,preprint]{aastex}
\newcommand{\nraoblurb}{The National Radio Astronomy Observatory is
a facility of the National Science Foundation operated under cooperative
agreement by Associated Universities, Inc.}
\newcommand{\cm}{$\,{\rm cm}$}
\newcommand{\m}{$\,{\rm m}$}

\newcommand{\au}{$\,{\rm AU}$}
\newcommand{\mas}{$\,{\rm mas}$}
\newcommand{\muas}{$\, {\rm \mu as}$}
\newcommand{\pc}{$\,{\rm pc}$}
\newcommand{\kpc}{$\,{\rm kpc}$}
\newcommand{\kms}{${\,{\rm km\, sec^{-1}}}$}

\newcommand{\s}{$\,{\rm s}$}

\newcommand{\mhz}{$\,{\rm MHz}$}

\newcommand{\hr}{$\,{\rm hr}$}
\newcommand{\cma}{$\,{\rm cm^{-3}}$}

\newcommand{\arcmper}{\rlap.{^{\prime}}}

\newcommand{\hi}{H~{$\scriptstyle {\rm I}$}}

\newcommand{\expo}[1]{${10^{#1}}$}

\begin{document}

\title{Limits on Turbulent \hi\ Fluctuations Towards PSR B0329+54
On Scales Between $0.0025$ and $12.5$\au }


\shortauthors{Minter et al. 2003}

\author{Anthony H. Minter\altaffilmark{1}, Dana
S. Balser\altaffilmark{1}, \& Jeyhan S. Kartaltepe\altaffilmark{2}$^{\rm ,}$\altaffilmark{3}}

\altaffiltext{1}{National Radio Astronomy Observatory,
P.O. Box 2, Green Bank WV 24944}
\altaffiltext{2}{Department of Physics and Astronomy, Colgate
University, Hamilton, NY 13346}
\altaffiltext{3}{currently at the University of Hawaii, Honolulu, HI 96822}

\begin{abstract}

We have measured the \hi\ absorption towards pulsar B0329+54 using the
Green Bank Telescope during 18 epochs between 2002 June 30 and 2003
October 10.  Three observing epochs consisted of a continuous period of 20
hours each, while fifteen epochs were 1--2 hours each.  
We calculate the structure function of
\hi\ absorption variations toward the pulsar on time scales of 10 minutes
to 16 months which, using the proper motion of 95\kms\ and the parallactic 
distance of $1.03$\kpc\ measured towards B0329+54 (Brisken et al. 2002), 
corresponds to angular scales of 0.37\muas\ to 23.8\mas\
and samples structures between $0.0025-12.5$\au\ assuming \hi\ gas half way 
to the pulsar and ignoring scintillation effects.  We find no
evidence for any turbulent \hi\ absorption fluctuations towards B0329+54, 
with the following upper limits on $\Delta\tau$ for various absorption 
features:
$0.026$ at
$-31$, $-21$, $-18$, and $+4$\kms\,
$0.12$ at $-11$\kms\, and $0.055$ at
$-1$\kms.

\end{abstract}


\section{Introduction}\label{sec:intro}

During the last decade, studies of \hi\ absorption lines have revealed
angular or temporal variations corresponding to spatial scales on the
order of tens of AU.  In some directions structure in the \hi\ has
been observed and interpreted as individual \hi\ clouds passing across
the line-of-sight with densities of \expo{4}--\expo{5}\cma.  Based on
these measurements a significant fraction (10--15\%) of the cold \hi\
gas must be in these clouds (Frail et al. 1994).  However, it is
difficult to reconcile the high \hi\ densities at AU-scales implied by
these measurements with other information about the interstellar
medium (ISM), as they would be greatly out of pressure equilibrium and
should be short lived.  Heiles (1997) proposed that the observed \hi\
components are formed from sheets and filaments where the large column
densities are produced by the appropriate viewing angle.  Deshpande
(2000) suggests that these data have been misinterpreted and that a
single power-law describes the distribution of cold \hi\ in the ISM.
It has also been suggested that the observed changes in the \hi\
absorption profiles towards pulsars are the result of scintillation
along with a velocity gradient in a uniform \hi\ medium
(Gwinn 2001).  Angular power spectra from \hi\ emission observations
reveal a power-law distribution of \hi\ structures on parsec scales
consistent with a turbulent medium (e.g., Green 1993).  It seems possible
that the AU-scale \hi\ fluctuations are part of the same turbulence
that is present on parsec scales, but this has not been conclusively
demonstrated.

If AU-scale fluctuations are present and are
part of the turbulent cascade seen at
parsec scales then it may be possible to determine some of the
fundamental characteristics of the \hi\ gas.  If the gas is purely
hydrodynamical (HD) then the smallest size scale that shows
fluctuations, called the inner scale, is larger than the ``molecular''
mean-free-path length (Tennekes \& Lumley 1994, Frisch 1996).  For
typical situations in the ISM the \hi\ mean-free-path length
corresponds to scales from $\sim 1-100$ AU.  
Collisions of the \hi\ with the ions and electrons in magneto-hydrodynamical
(MHD) turbulence
could create \hi\ fluctuations on scales smaller than $\sim $ AU.
If we are able to measure the inner
scale of the HD turbulence then it is possible to estimate the
kinematic viscosity of the \hi\ in the ISM.  
The kinematic viscosity
is given approximately by ${\rm \nu \sim \lambda_{mfp} c_{th}}$, where
${\rm \lambda_{mfp}}$ is the mean-free-path length and ${\rm c_{th}}$
is the thermal sound speed.  
If the \hi\
turbulence is MHD then sub-AU scale \hi\ fluctuations would be present
which would tell us that magnetic fields cannot be ignored
in any aspect of the dynamics of interstellar gas.

We have made \hi\ absorption measurements towards the pulsar B0329+54
with the Green Bank Telescope (GBT) to measure \hi\ fluctuations on
sub-AU scales.  B0329+54 has a parallactic distance of 
$ 1.03^{+0.13}_{-0.12}$\kpc\ and a proper motion of 
$ 95^{+12}_{-11}$\kms\ (Brisken et
al. 2002).  The pulsar's proper motion corresponds to 
scales of 0.03\au\ in one day and 1\au\ in one month for an \hi\ cloud 
half way to the pulsar.

\section{Observations}\label{sec:obs}

The 100\m\ Green Bank telescope (GBT) of the National Radio Astronomy
Observatory\footnote{\nraoblurb} (NRAO) was used for this measurement.
The GBT is an unblocked aperture telescope with a spatial resolution
of 9$\arcmper$2 at 21\cm.  The system temperature on cold sky was
$\sim 20$ K.  The detector was the NRAO spectral processor, an FFT
spectrometer, configured to have 1024 channels for each linear
polarization with a bandwidth of 1.25\mhz, producing a spectral
resolution of 0.26\kms\ per channel.  The accumulation memory is
32-bit, providing good dynamic range.  The spectral processor
integration time was 14 pulse periods, approximately 10 seconds.

The pulsar B0329+54 is an ideal target for study of small-scale
structure in cold \hi.  It is very bright so a change in opacity of 0.1
can be detected in a time less than the scintillation time-scale of
$\sim 15$ minutes.  Its
declination is such that it can be observed by the GBT continuously
for $\sim 20$\hr\ 
during which $\sim 50$ scintles are seen.  The
observations reported here were made in 18 separate observing
sessions.  Three long ($\sim 20$\hr) sessions, separated by two weeks,
were made to probe \hi\ fluctuations on sub-AU scales.  Fifteen short
($\sim 1-2$\hr) observations spread over the subsequent
fifteen months sample scales larger than an AU.

The data were calibrated using a method similar to that described by
Weisberg (1978).
For each integration an absorption spectrum was formed by taking
the difference between the pulsar ``on'' and pulsar ``off'' spectra.
If $\tau_{\rm i,j}$ is the \hi\ opacity for
the i$^{\rm th}$ spectral channel and the j$^{\rm th}$ integration
sample then
\begin{equation}
T_{\rm i,j}(\rm p_{\rm on})~ e^{-\tau_{\rm i,j}}
= \frac{ T_{\rm i,j}({\rm p_{\rm on}, \ell_{on}}) -
T_{\rm i,j}({\rm p_{\rm off}, \ell_{\rm on}}) }{ T_{\rm i,j}({\rm
p_{\rm off}, \ell_{\rm off}}) / 
\sum_{i=1}^{n_{\rm chan}} T_{\rm i,j}({\rm p_{\rm off}, \ell_{\rm off}}) }
\end{equation}
where $n_{\rm chan}$ is the number of spectral channels.
The symbol $T$ denotes intensity in units of Kelvins where, for
example, $T_{\rm i,j}({\rm p_{\rm off}, \ell_{on}})$ corresponds to the
intensity at the frequency of the \hi\ line emission when the pulsar is 
``off'', $T_{\rm i,j}({\rm p_{\rm off}, \ell_{off}})$ corresponds to the
intensity at frequencies other than the \hi\ line emission when the pulsar is 
``off'',
and  $T_{\rm i,j}(\rm p_{\rm on})$ corresponds to the intensity at
all observed frequencies when the pulsar is ``on''.
The power is converted
from detector counts to Kelvin by using a calibrated noise diode that
was injected every pulsar cycle for 10\% of the pulsar period.
Each integration is then weighted and summed such that
\begin{equation}
\left<T_{\rm i}(\rm p_{\rm on})\right> e^{-\tau_{\rm i}} = 
\frac{\sum_{j=1}^{n_{\rm int}} T_{\rm i,j}(\rm p_{\rm on})~
e^{-\tau_{\rm i,j}} \frac{
T^2_{\rm i,j}({\rm p_{\rm on}, \ell_{off}})}{T^2_{sys_{\rm i,j}}}}
{\sum_{j=1}^{n_{\rm int}} \frac{
T^2_{\rm i,j}({\rm p_{\rm on}, \ell_{off}})}{T^2_{sys_{\rm i,j}}}}
\label{eqn:tpe}
\end{equation}
where $T_{sys_{\rm i,j}}$ is the system temperature,
$ \left<T_{\rm i}(\rm p_{\rm on})\right>$ is the weighted average pulsar flux, 
and there are
$n_{\rm int}$ independent integration samples.

The flux from B0329+54 varies for two reasons: 1) pulse to pulse variations
due to intrinsic emission variations; and 2) due to scintillation.
Because B0329+54 varies in intensity during our 10\s\ integration
period and can be a significant fraction of the total system
temperature, some of the emission spectrum will appear in the
absorption spectrum as ``ghosts''.  Weisberg (1978) has shown that the
measured pulsar absorption spectrum is a linear combination of $T_{\rm
i}({\rm p_{\rm off}, \ell_{on}})$ and the desired pulsar absorption
spectrum.  Therefore the measured \hi\ emission spectrum is used to
fit and remove the observed ``ghosts''.  Simultaneously a polynomial
model is used to remove the average pulsar flux 
($\left<T_i(\rm p_{\rm on})\right>$),
which cannot be independently measured, 
and any structure in the baseline due to instrumental effects.
The ``ghost'' and polynomial fit is constrained only in spectral regions 
where there is no \hi\ absorption.  The ``ghost'' spectrum and the
polynomial fit is then extrapolated through the regions with
\hi\ absorption.
Since the 1.25~MHz bandwidth of these observations is a significant fraction 
of the scintillation bandwidth\footnote{The scintillation bandwidth is the
$e^{-1}$ scale over which the pulsar's signal becomes decorrelated in
frequency due to scintillation of the pulsar signal from a non-uniform
distribution of electrons in the ISM.  It effectively represents the average
frequency domain size of an observed scintle.}   for B0329+54
(measured to be $5.1\pm0.1$~MHz at $1640$~MHz by Minter (2001) ), 
the average pulsar flux 
contains scintillation induced structures that vary slowly with
frequency.  In order to remove the baseline structure arising from 
scintillation structures and instrumental effects, a 5th order polynomial 
is used.

For each integration the RMS noise was computed for the
part of the spectrum without \hi\ line emission for both
$T_{\rm j}(\rm p_{\rm on})$ and  $T_{\rm j}(\rm p_{\rm off})$.
Since the \hi\ line emission contributes significantly to the system 
temperature ($\sim 20~{\rm K}$) the RMS noise will be larger 
for frequencies where there is \hi\ line emission.  Thus the 
frequency dependent noise is given by
\begin{equation}
\sigma_{T_{\rm i,j}(\rm p_{\rm on})} = 
\sigma_{T_{\rm j}(\rm p_{\rm on})} \left( 1 + T^{\rm HI}_{\rm i}/T_{sys} 
\right)
\end{equation}
where $T^{\rm HI}_{\rm i}$ is the \hi\ line emission strength in the
ith spectral channel.
A similar expression is used for $\sigma_{T_{\rm i,j}(\rm p_{\rm off})}$.
These errors are then propagated for all subsequent calculations.

Systematic errors are also important in the comparison of \hi\ absorption
towards pulsars from different epochs.  The systematic
errors arise from the
removal of the average pulsar flux, $\left<T_i(\rm p_{\rm on})\right>$, 
via a polynomial fit from the results of equation~\ref{eqn:tpe} in order
to determine the $e^{-\tau}$ spectra.  The actual pulsar flux will
certainly not follow the value of the polynomial as it is extrapolated
over the regions of \hi\ absorption that were excluded from the polynomial
fitting.  This difference between the actual pulsar flux and the extrapolated
polynomial value results in a systematic error that must be taken into
account when comparing the \hi\ absorption spectra between different epochs.

\begin{figure}
\epsscale{0.9}
\includegraphics[scale=0.6,angle=-90]{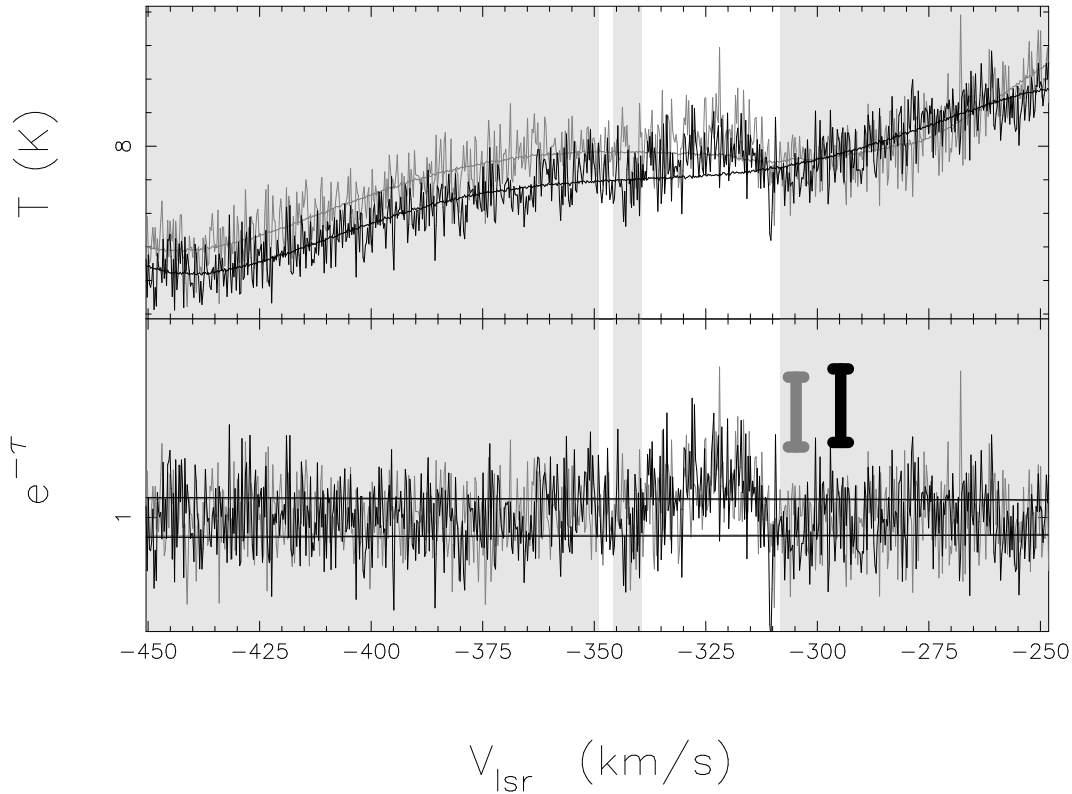}
\caption{Top panel: A spectrum of $\left<T_{\rm i}(\rm p_{\rm on})\right>
e^{-\tau_{\rm i}}$ for two polarization (X and Y) towards 0329+54 and 
offset in
frequency from the \hi\ line.  The smooth, solid curves are the
polynomial fits to the data.  
The grey area denotes the channels used in the polynomial fit where as the
white area is where the polynomial was extrapolated.  The same regions
(by channel number) used here were also used for the \hi\ line data.
Bottom Panel: The resulting spectrum of $e^{-\tau_{\rm i}}$.  The 
``straight lines'' indicate the 
$1\sigma$ level of the random noise and
the ``thick'' error bars indicate the 
$1\sigma$ errors 
in extrapolating 
the polynomial.}
\label{fig:noise}
\end{figure}

In Figure~\ref{fig:noise} we show an example of how the systematic error
is calculated.  A frequency band without \hi\ emission or absorption was 
observed and reduced in the same manner as the pulsar
\hi\ absorption data.  The upper panel of Figure~\ref{fig:noise}
shows the results of equation~\ref{eqn:tpe} and the polynomial fit.  The
polynomial fit has only been done in the shaded regions, which are the
same channel numbers used for the polynomial fit for the \hi\ absorption 
measurements.
The bottom panel of Figure~\ref{fig:noise} shows the resulting $e^{-\tau}$
spectrum.  As can been seen in Figure~\ref{fig:noise}, there is excess
noise in the regions where the polynomial fit was extrapolated 
(the white areas in  Figure~\ref{fig:noise}).  This
systematic noise is shown by the single error bar above the data in the
bottom panel of Figure~\ref{fig:noise}.  The random noise is shown by
the two solid lines in the bottom panel of Figure~\ref{fig:noise}.
The systematic noise was determined by taking the mean deviation
of the $e^{-\tau}$ spectra away from $e^{-\tau}=1$ within the region 
where the polynomial fit was extrapolated (the white areas in the bottom
panel of Figure~\ref{fig:noise}) and then subtracting, in quadrature,
the random noise term.
The systematic noise was determined by (i) subtracting the polynomial
obtained by omitting the \hi\ absorption channels (shown in
Figure~\ref{fig:noise}) with a polynomial obtained by using all of
the channels; and (ii) taking the rms of the difference between these
two fitted spectra over the omitted channel ranges.

The average systematic error found using this method amounts to $\pm 0.005$ 
in the $e^{-\tau}$ spectra.  
Given that previous detections of \hi\ fluctuations have
been recently questioned (e.g. Johnston et al. 2003, 
Stanimirovi\'{c} et al. 2003) we choose to be conservative in our
error estimates.
The quoted uncertainties for our measurements thus contain both random and
systematic errors, which have been added in quadrature. 
We use a value of $\pm 0.005$ in the $e^{-\tau}$ spectra, as determined
above, for the systematic errors in all \hi\ spectra.

\section{Results}\label{sec:results}

\begin{figure}
\epsscale{0.9}
\includegraphics[scale=0.6,angle=-90]{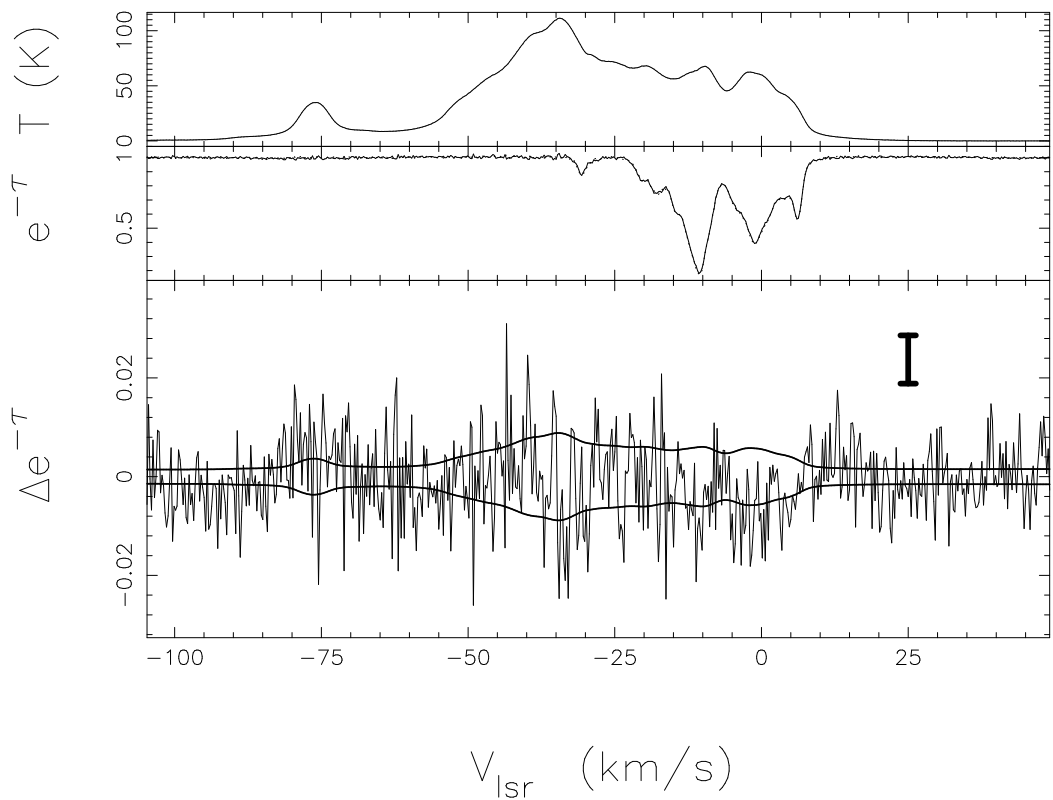}
\caption{Top: The \hi\ emission spectrum towards B0329+54.  Middle: 
The absorption spectra from July 14, 2002 (solid line) and July 2, 2003
(dotted line).  Bottom: Difference between the two absorption spectra.  The
lines above and below the spectrum are the $1\sigma$ uncertainties.
The uncertainty is a function of frequency since the \hi\ emission 
contributes significantly to the total system temperature.  The 
$1\sigma$
uncertainty in extrapolating the polynomial fit to the pulsar emission
spectrum across the \hi\ line is shown by the individual error bar.}
\label{fig:spectrum}
\end{figure}

Figure~\ref{fig:spectrum} compares the absorption spectrum towards
B0329+54 for two epochs, July 14, 2002 and July 2, 2003.  
The solid lines above and below the difference spectrum are the $1 \sigma$
random noise uncertainties.  The systematic error from the polynomial
fitting is indicated by the error bar above the data.
There are no
significant differences between the two spectra shown in 
Figure~\ref{fig:spectrum}.
In fact, no variations have been detected over the entire 16 months of
observations.
  
The structure function for time variations in the opacity is defined as
\begin{equation}
D_{\tau}(\Delta t,i) = \left< \left( \tau(t,i) - 
\tau(t+\Delta t,i) \right)^2 \right>
\end{equation} 
where $\tau(t,i)$ is the opacity at time $t$ in the ith spectral channel.
The structure function is actually computed by binning values of
$\left( \tau(t,i) - \tau(t+\Delta t,i) \right)^2$ and then averaging
the data within a given bin.  The errors for the structure function are
computed from the distribution of values within a given bin.
Analysis of the
data using a structure function has two advantages: (1) it reduces
the noise for a given delay through averaging; and (2) it provides
information on the power spectrum of \hi\ absorption fluctuations.
If the power spectrum of opacity variations is a power law,
as is the case for turbulent fluctuations, then the 
structure function will also be a power law.  If only noise is present in the
data then the structure function will be constant at a level of 
$2 \sigma^2_{\tau(i)}$.   (We refer the reader to
 Spangler et al. (1989) for a more detailed discussion of structure functions.)

\begin{figure}
\epsscale{0.75}
\includegraphics[scale=0.35,angle=-90]{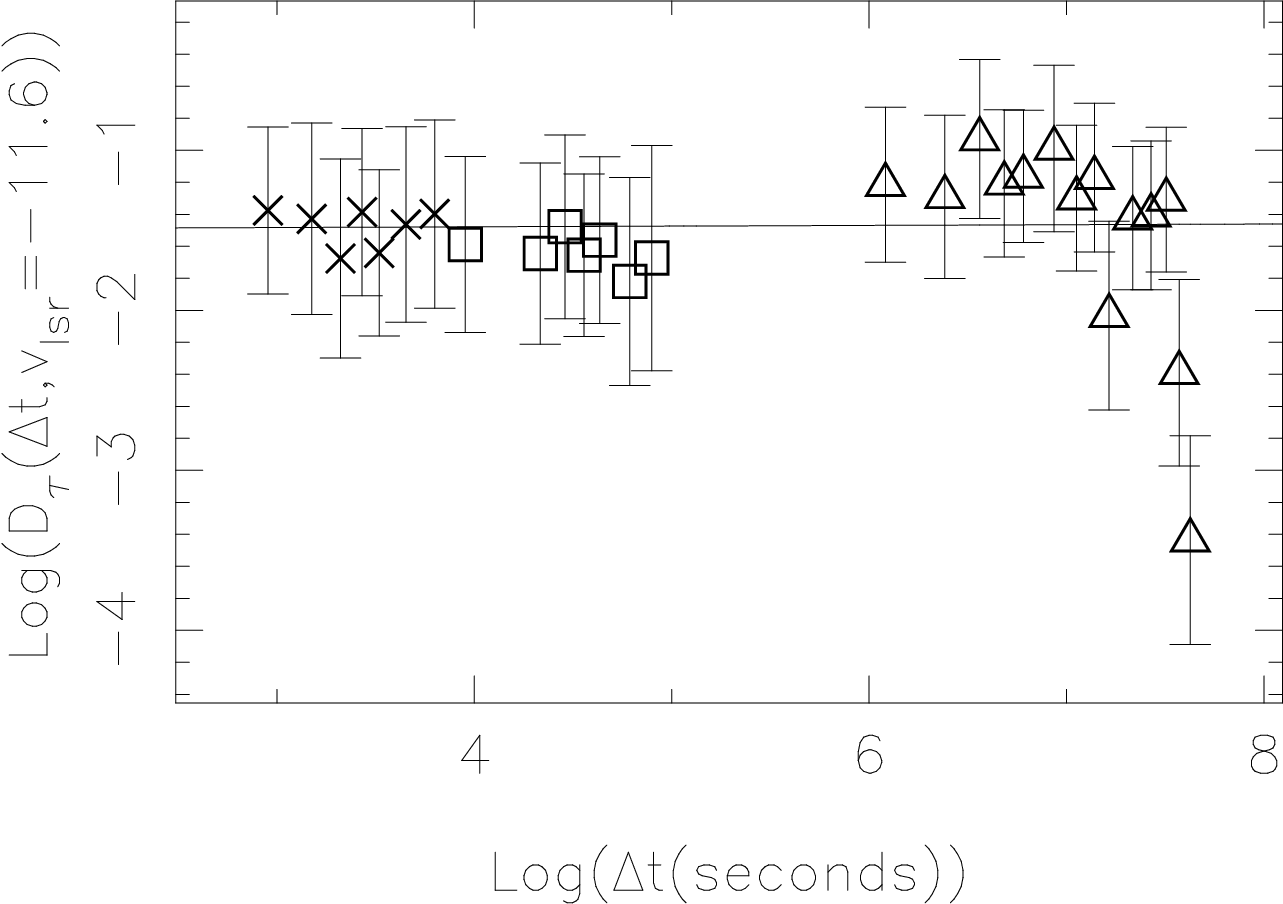}
\includegraphics[scale=0.35,angle=-90]{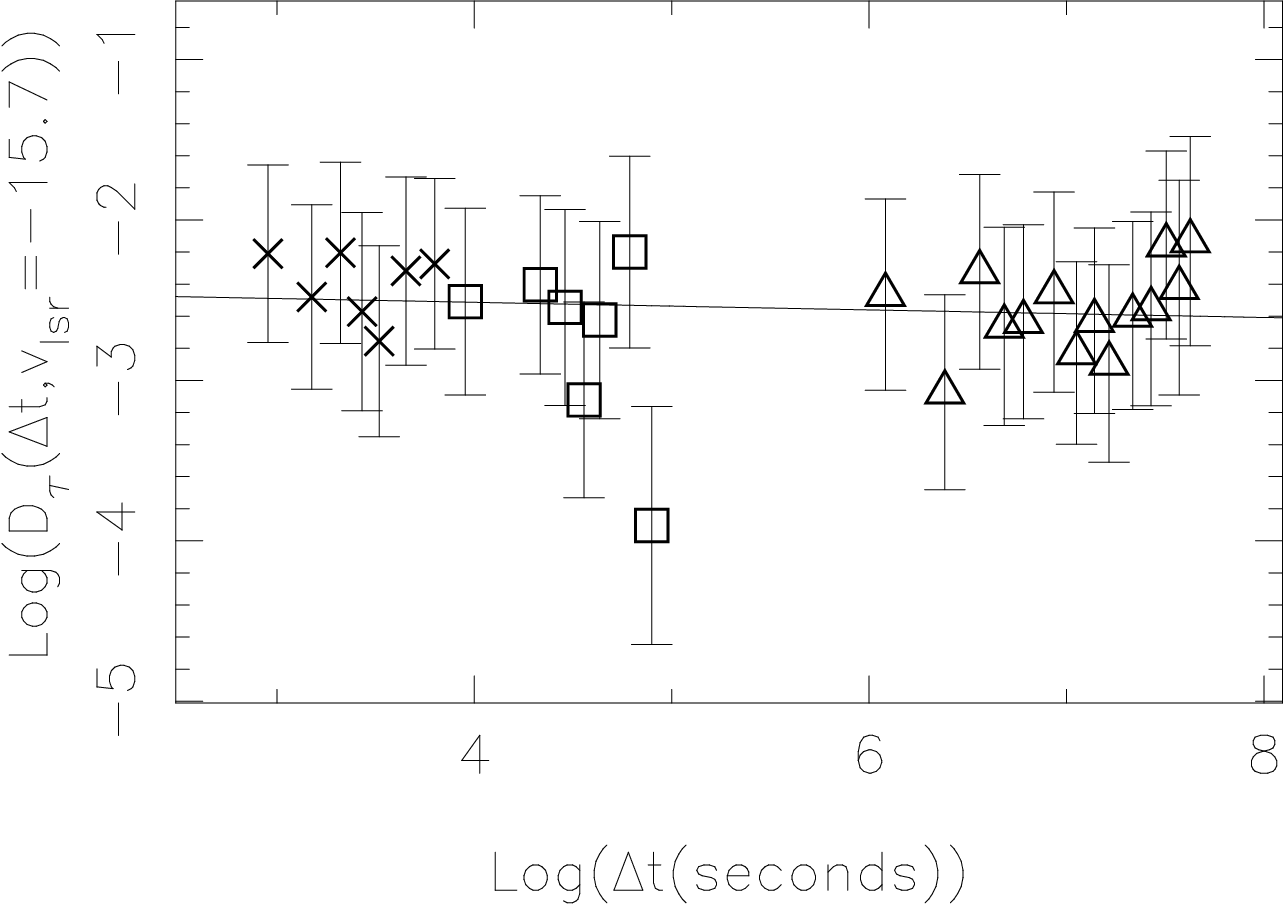}
\caption{Structure function of $\tau$ for the spectral channels
centered on ${\rm v_{LSR} = -11.6 ~km/s}$
(left panel), at the peak of the absorption line,
and ${\rm v_{LSR} = -15.7 ~km/s}$ (right panel), halfway down the
absorption line,
with 
$3 \sigma$ error bars. 
The cross symbols are data averaged over a
single scintle, the square symbols consist of $\sim 2$\hr\ averages
within a given epoch, and the triangle symbols are data averaged over
a single observing epoch.
The best fit power laws  are shown as the solid lines and are given by  
$(-1.5 \pm 0.1) + (0.00 \pm 0.02) \cdot \log(\Delta t)$
 and $(-2.4 \pm 0.1) + (-0.02 \pm 0.02) \cdot \log(\Delta t)$
respectively.  
No turbulent \hi\ opacity fluctuations are detected on any scale between $0.0025$ and
$12.5$\au.}
\label{fig:sf}
\end{figure}

\begin{figure}
\epsscale{0.75}
\includegraphics[scale=0.6,angle=-90]{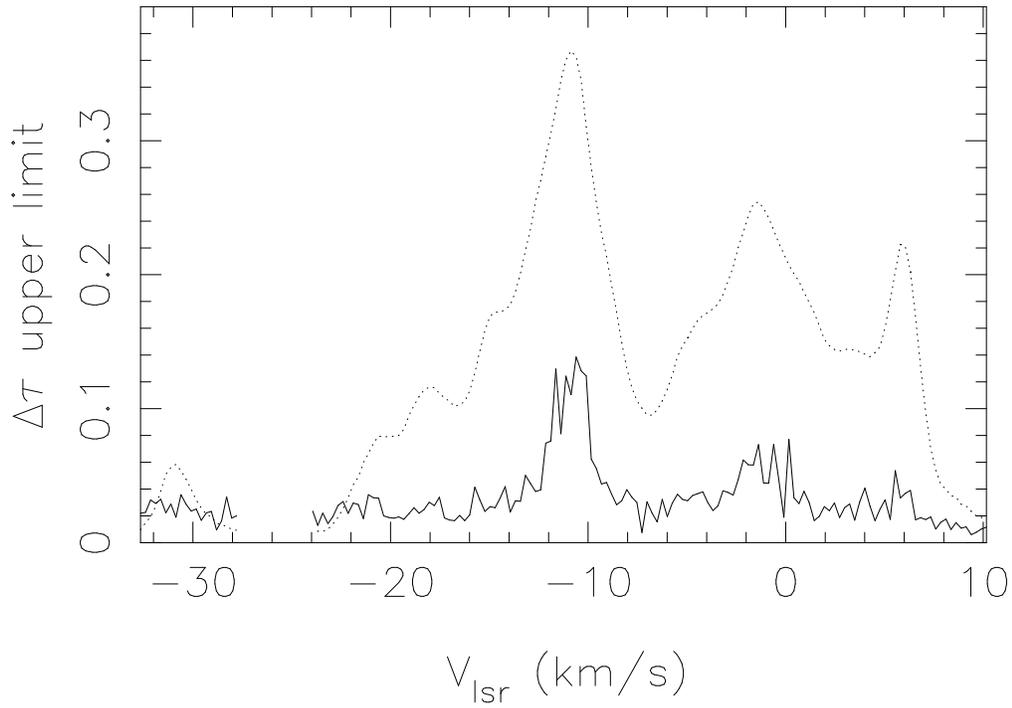}
\caption{Upper limits to $\Delta \tau$ from our measurements (solid line)
and from the scintillation model of Gwinn (2001) (dotted line).  The 
frequency dependence of the $\Delta \tau$ upper limits follows the expected 
noise contributions due to the change in system temperature from the 
\hi\ emission and absorption.
}
\label{fig:taulimit}
\end{figure}

Figure~\ref{fig:sf} plots the structure function for our observations
of B0329+54 for two frequency channels. 
The structure function data
are consistent with there being no turbulence in the \hi\ absorption
down to our detection limits.
Figure~\ref{fig:taulimit} shows our
observational upper limits to the turbulent fluctuations 
in opacity derived from the
structure functions in each spectral channel.
The data cover differences between ten minutes to sixteen months
corresponding to angular scales of 0.37\muas\ to 23.8\mas\
and geometric linear size scales of $0.0025-12.5$\au, assuming \hi\ gas half 
way to the pulsar and using the pulsar's proper motion velocity of 95\kms\
and distance of 1.03\kpc.
(For the remainder of this paper we calculate 
 distance scales
assuming a distance halfway to the pulsar of $515$\pc.)
Furthermore, comparing our spectra to
those of Gordon et al. (1969), who first detected \hi\ absorption
towards B0329+54, shows no evidence for variations in \hi\ absorption
greater than $\Delta e^{-\tau} \sim 0.01$ (the noise level of the
Gordon et al. measurements is $\Delta e^{-\tau} \sim 0.01$ while
the noise for our measurements is $\Delta e^{-\tau} \sim 0.002$) 
with a velocity resolution of 1.7\kms.
This corresponds to a linear scale of $\sim 350$\au.

Strong interstellar scintillation is observed towards B0329+54.  
The multi-path propagation that results from scintillation 
effectively creates a spatial smoothing of \hi\ absorption 
structures within the observed angularly broadened size of the pulsar
which might mask absorption variations that exist on very small scales.  
Semenkov et al. (2003) have limited
the angular broadening size of B0329+54 at 1600\mhz\ to be $\leq
1.8$\mas\ corresponding to a scintillation smoothing scale of $\leq 1.15$\au\
assuming a $\lambda^2$ scaling of the angular
broadening size.  Our observations probe scales up to $12.5$\au\ 
and thus scintillation cannot account for a
lack of \hi\ absorption fluctuations in the data.  
The scintillation smoothing provides a lower limit 
to the smallest scale opacity structure that we can probe.
However we only have upper limits to the scattering disk size.
Therefore we use the geometric linear size scale of 0.0025\au\ (see above)
rather than the scattering
disk size as the smallest size scale probed by our observations. 

Gwinn (2001) suggests that interstellar scintillation coupled with
gradients in the Doppler velocity of \hi\ can produce small-scale
fluctuations in \hi\ absorption spectra towards pulsars.  The GBT
observations of B0329+54 can decouple density fluctuations from
velocity gradients since an absorption spectrum can be measured for
individual scintles.  The data points in Figure~\ref{fig:sf} are
divided into three groups.  The cross symbols are data averaged over a
single scintle, the square symbols consist of $\sim 2$\hr\ averages
within a given epoch, and the triangle symbols are data averaged over
a single observing epoch.  No turbulent fluctuations are detected
on any timescales from the scintillation timescale to sixteen months
in our observations.  

\begin{figure}
\epsscale{0.75}
\includegraphics[scale=0.6,angle=-90]{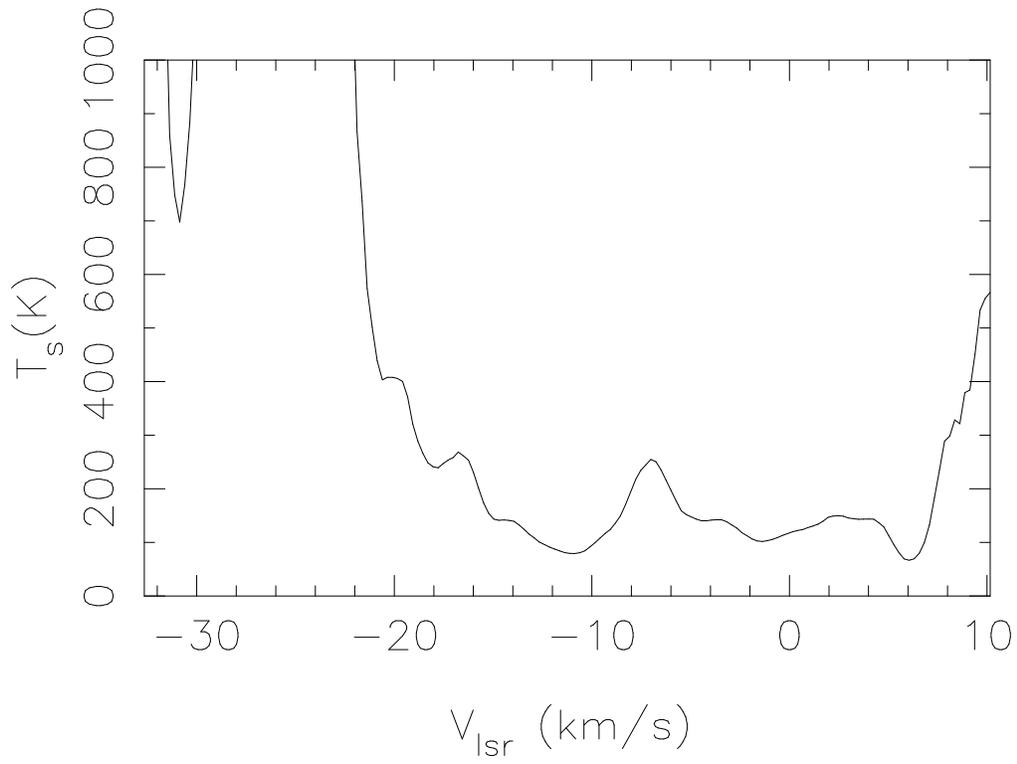}
\caption{Upper limits to the spin temperature, ${\rm T_s}$, of the \hi\
responsible for the absorption towards B0329+54.}
\label{fig:tspin}
\end{figure}

Gwinn (2001) predicts that the opacity variations that would be
observed are given by $\sigma_\tau = \sqrt{ {\tau \Delta v \over c_s}}$
where $\Delta v$ is the amount by which the velocity of the absorption
feature changes and $c_s$ is the thermal sound speed.
We have fit Gaussians to the peak absorption features for each epoch.
From these fits we find no trend in the value of the line
center and can limit any change in velocity of the absorbing gas to 
$\Delta v < 0.065$\kms.
We can derive upper limits
to the thermal sound speed from the \hi\ spin temperature (${\rm T_s}$)
found from comparing the \hi\ emission spectrum (${\rm T_{em}}$) 
with the \hi\ absorption spectrum.
The spin temperature is found from ${\rm T_s = {T_{em}\over 1 - 
e^{-\tau}} }$ and should be considered an upper limit since some
warm \hi\ contributes to ${\rm T_{em}}$ but typically does not contribute
to the \hi\ absorption.  The spin temperature upper limits are shown
in Figure~\ref{fig:tspin}.  The sound speed can then be estimated from
${c_s = {\rm 0.093 \sqrt{T_s(K)}}}$\kms. 
An estimate of Gwinn's $\sigma_\tau$ prediction is shown as the dotted line 
in Figure~\ref{fig:taulimit}.  
The derived spin temperatures are not likely to be incorrect by more
than a factor of a few.  We thus consider the predictions using Gwinn's
formula for opacity variations shown in Figure~\ref{fig:taulimit} to be
a reasonable upper limit for opacity variations induced by the combination
of scintillation and velocity gradients.  From Figure~\ref{fig:taulimit}
it can be seen that our measured limits are smaller than the upper limits
determined for Gwinn's $\sigma_\tau$ formula.  Although this suggests
that Gwinn's hypothesis could be incorrect, we cannot claim this with
any certainty since only an upper limit can be determined 
for Gwinn's prediction.


\section{Discussion}\label{sec:discussion}

Our GBT observations towards B0329+54 were made primarily to probe
sub-AU \hi\ structures.  The structure functions of the change in opacity
versus time are consistent with noise (i.e. the
structure functions are constant values).  The structure functions
thus place upper limits on the opacity variations
as discussed in \S~3.  
So we do not detect \hi\ opacity
variations consistent with a turbulent power law distribution
on scales $< 12.5$\au\ greater than $0.026$ for 
the $-31$, $-21$, $-18$, and $+4$\kms\ absorption lines,
$0.12$ for the $-11$\kms\ absorption line, and $0.055$ for 
the $-1$\kms\ absorption line.  This is somewhat 
surprising since Frail et al. (1994) detected
\hi\ variations of $\Delta \tau \sim 0.1$ in all observed pulsars.
However, Johnston et al. (2003) have made
multi-epoch observations of \hi\ absorption towards four southern
pulsars and find no significant variations and
Stanimirovi\'{c} et al. (2003) reach a similar conclusion from
re-observations of the Frail et
al. pulsars to increase the number of temporal baselines.  
VLBA \hi\ absorption measurements have been summarized by Faison (2002).
Only two sources show significant \hi\ fluctuations.  
Small-scale \hi\ structure measured through
\hi\ absorption is currently found in only two of the
15 sources observed in Johnston et al. (2003), Stanimirovi\'{c} et al. (2003),
Faison (2002) and this work.

On large scales the distribution of \hi\ is influenced by spiral
density waves, supernovae, etc. as is evident by the observed shells
and filaments in neutral hydrogen surveys.  Dickey \& Lockman (1990)
have argued that no more than 10\% of the total \hi\ exists in
small-scale structures ($\le\ 1$\pc).  That is, there are not large
variations in the \hi\ column density on small spatial scales and the
concept of distinct \hi\ clouds does not describe most of the Galactic
neutral hydrogen.  The distribution of \hi\ can be analyzed by
producing angular power spectra of \hi\ emission over different
spatial scales.  The results are well fit by a power-law with a slope
of approximately $-3$ towards different directions in the Galaxy
(e.g., Crovisier \& Dickey 1983; Green 1993; Dickey et al. 2001).
These results can be described by a turbulent cascade of energies
(Lazarian \& Pogosyan 2000).  But does this turbulence extend down to
very small spatial scales?  In other words, what is the inner scale of
the turbulence?  

We are unable to answer this question with the current data on
B0329+54 since we have not detected any significant 
turbulent \hi\ fluctuations.

Of particular note are the results of Shishov et al. (2003) from 
diffractive scintillation measurements of B0329+54.  They find that
on scales below $3 \times 10^{16}~\rm cm$ that the diffractive scintillation
can be explained with a scattering screen comprised solely of ionized 
gas.\footnote{The value $3 \times 10^{16}~\rm cm$ was obtained by 
Shishov et al. (2003) using a velocity of $139$\kms\ for
the pulsar.  Using a velocity of $95$\kms\ as measured by Brisken et al. (2002)
results in a scale of $2 \times 10^{16}~\rm cm$.} However,
on scales above  $3 \times 10^{16}~\rm cm$ their results require that some 
neutral gas is also present within the scattering screen.  
Could this indicate that the inner scale for the neutral gas is $2000$~AU?  
If this is the inner scale then an \hi\ density of $1~\rm cm^{-3}$ at a 
temperature of $100~\rm K$ would give a kinematic viscosity of 
$3 \times 10^{21}~\rm cm^2 ~s^{-1}$.

\acknowledgements

The research of J.S.K. at NRAO was supported by the NSF Research
Experiences for Undergraduates program.
We thank Crystal Brogan, Avinash Deshpande, and Sne\'{z}ana
Stanimirovi\'{c} for stimulating discussions.  We thank Jay Lockman
for many discussions about the observations and calibration procedures
and for commenting on the manuscript.
We would like to thank the anonymous referee whose comments greatly
improved the quality of this work.

\clearpage


\end{document}